\documentclass[amsfonts,aps, prl,nofootinbib, twocolumn,showpacs, showkeys,longbibliography]{revtex4-1}

\usepackage{graphicx} 
\usepackage{amsmath}

\newcommand{\D}{\ensuremath\mathrm{d}}

\begin{document}

	\author{Stefan Thurner$^{1,2,3*}$ and Benedikt Fuchs$^1$}

	\affiliation{$^1$Section for Science of Complex Systems, Medical University of Vienna, Spitalgasse 23, A-1090, Austria\\ 
	$^2$Santa Fe Institute, 1399 Hyde Park Road, Santa Fe, NM 87501, USA\\
	$^3$International Institute for Applied Systems Analysis, Schlossplatz 1, A-2361 Laxenburg, Austria}
	\email{stefan.thurner@meduniwien.ac.at}
	
	\title{Physical forces between humans and how humans attract and repel each other based on their social interactions in an online world} 
	
	\date{\today}

\begin{abstract}
{Physical interactions between particles are the result of the exchange of gauge bosons. 
Human interactions are mediated by the exchange of messages, goods, money, promises, hostilities, etc. 
While in the physical world interactions and their associated forces have immediate dynamical consequences 
(Newton's law) the situation is not clear for human interactions. Here we study the acceleration between humans who 
interact through the exchange of messages, goods and hostilities in a massive multiplayer online game. For this game we have 
complete information about all interactions (exchange events) between about 1/2 million players, and about 
their trajectories (positions) in a metric space of the game universe at any point in time. 
We derive the interaction potentials for communication, trade and attacks and show that they are harmonic in nature. 
Individuals who exchange messages and trade goods generally attract each other and start to separate immediately after exchange events stop. 
The interaction potential for attacks mirrors the usual ``hit-and-run'' tactics of aggressive players.
By measuring interaction intensities as a function of distance, velocity and acceleration, we show that ``forces'' between players 
are directly related to the number of exchange events. The power-law of the likelihood for interactions vs. distance   
is in accordance with previous real world empirical work. 
We show that the obtained potentials can be understood with a simple model assuming an exchange-driven 
force in combination with a distance dependent exchange rate.}
\end{abstract} 


\maketitle

\section{Introduction}

\begin{figure*}
	\centering
	\includegraphics[width=\textwidth]{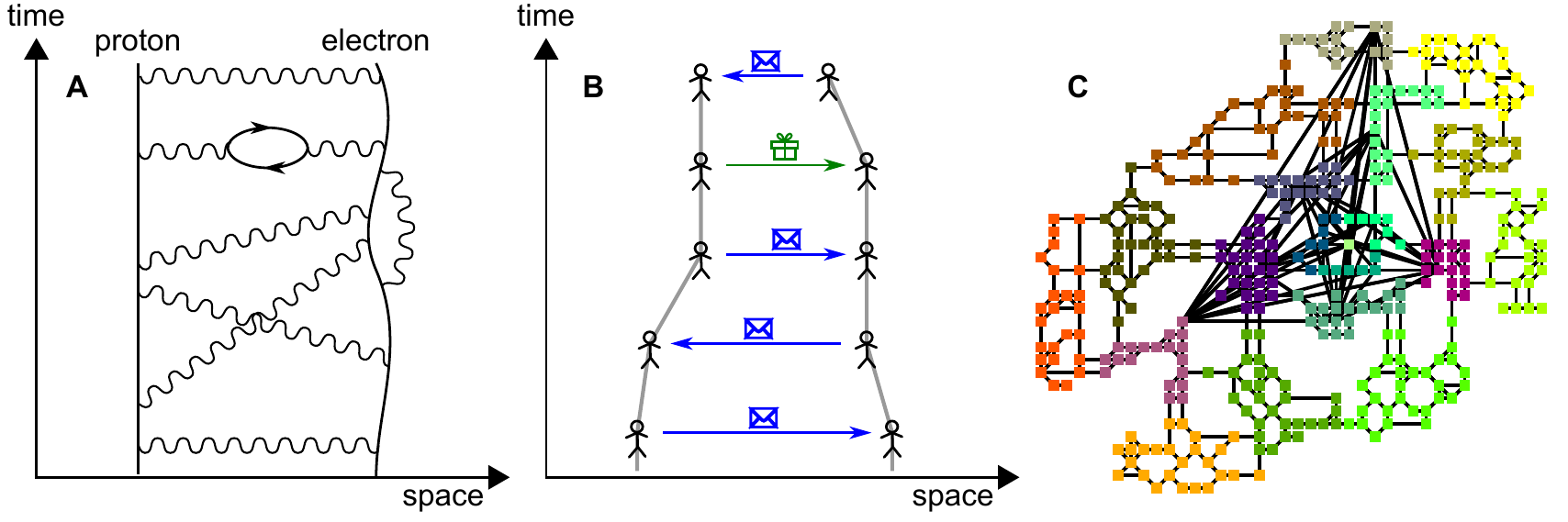}
	\caption{Interactions mediated by exchange of particles.
	A Interaction between a proton and an electron by exchange of virtual photons.
	 (after \cite{Feynman1985}). 
	B Two players in \emph{Pardus} interact by exchanging messages and goods.
	C Map of the \emph{Pardus} universe. Nodes are sectors (cities), lines between them are connections (streets).
	Colors represent different regions (countries) \cite{Szell2012}.}
	\label{fig:Illustration}
\end{figure*}

The four forces of physics can be understood in terms of an exchange of virtual bosons between interacting particles. 
Electromagnetism exchanges photons, see Fig.~\ref{fig:Illustration} A, the weak force W- and Z-bosons, and  
the strong force gluons. Gravitation is thought to be mediated by the exchange of (hypothetical) gravitons.
By treating virtual exchange particles as excitations of a field the functional form of the interaction potential can be derived
\cite{Feynman1998,Wilson1974}.

Ignoring quantum field theory for the remainder of this paper, given the potential 
$V_{}(\mathbf{x})$ implies Newton's law,
\begin{equation}\label{eq:Newton}
m\frac {\D^2} {\D t^2} \mathbf{x} = -\nabla V_{}(\mathbf{x}) \quad ,
\end{equation}
which is usually rephrased as a one particle problem with a central force, $V_{}(\mathbf{x})=V_{}(r)$,
where $r$ is the distance between the two particles, and we have
\begin{equation}\label{eq:NewtonRadial}
	ma = -\frac{\D}{\D r}\left[V_{}(r)+V^0(r) \right] \quad , 
\end{equation}
where $V^0(r)$ corresponds to a pseudo-potential, which could be caused for example by the angular momentum 
$V^0(r) = \frac{L^2}{2mr^2}$ (in cylindrical coordinates).

Similar to physical interactions, human interactions are to a large extent based on the concept of exchange. 
The objects exchanged can be information, messages, goods, money, presents, promises, 
aggression (e.g. bullets), etc. In Fig.~\ref{fig:Illustration} B we schematically draw the trajectories of two individuals who exchange  
messages and a gift; their relative distance reduces over time.

It is in general not possible to experimentally verify if exchange events between humans generate effective attractive or repulsive 
forces. This is due to the lack of simultaneous information of exchange events and the trajectory of humans.   
The existence of potentials causing and influencing the relative motion of humans is not new and has been conjectured in \cite{Helbing1995}.
New technologies in data acquisition and storage are about to change the experimental situation. 
Data from mobile phone networks, email networks, and several online social network sites show that the probability for interaction events 
decays as an approximate power-law with distance, $P\propto r^{-\alpha}$ \cite{Lambiotte2008,Krings2009,Levy2014,Liben-Nowell2005,Adamic2005,Backstrom2010,Scellato2011,Cho2011,Grabowicz2014}, with exponents ranging from  
 $\alpha=0.83$ \cite{Cho2011} to $\alpha=2.0$ \cite{Lambiotte2008,Krings2009,Levy2014}, see Tab.~\ref{tab:exponents}.
Few empirical studies go beyond the analysis of the relation between distance and social dynamics.  
It was found  that humans mostly travel towards others with whom they share a (weak) tie in \cite{Phithakkitnukoon2012}. 
In \cite{Cho2011} human mobility is described as a combination of a periodic daily pattern (from ``home''  to  ``work'') 
and long-distance travels which are influenced by social networks. This model was successfully applied to mobile phone data 
and the social networks {\em Brightkite} and {\em Gowalla} \cite{Cho2011}. 

In this work we study a unique data set containing all interactions between the players of the massive multiplayer online game (MMOG) \emph{Pardus}.
At the same time we know the players' exact positions at any point in time. 
The MMOG has been extensively studied as a human model society 
\cite{Szell2010msd,Szell2010mol,Szell2012,Thurner2012,Szell2013,Klimek2013,Corominas-Murtra2014,Sinatra2014,Fuchs2014a,Fuchs2014b}.
\emph{Pardus} has more than 430,000 players who ``live'' in a virtual environment and interact with each other in a multitude of ways.
The game is open ended and players pursue their self-defined goals.
Players earn virtual currency through economic activities such as mining raw materials, processing them, or trading.
We consider trading between two players as one form of an exchange event; it usually involves the exchange of goods against currency.  
Players communicate with each other through the exchange of messages via an internal messaging system which is comparable to one-to-one emails.
There exist destructive forms of interaction where players attack each other if they are in close proximity. 
Additional types of interaction which are not considered further in this work include friendship and enmity markings, 
destruction of equipment, revenge, piracy, and indirect interactions through the formation of groups and governance. 
For further details see e.g. \cite{Szell2010mol,PardusHP}. 
Note that communication can happen over large distances, trading (exchange of goods) and attacks (exchange of bullets) require temporal 
and spatial ``locality''.  

The game is constrained to a 2 dimensional virtual universe that is partitioned into 400 so-called ``sectors'' (which play the role of cities). Sectors 
are connected by 1064 local and 77 long-range connections (streets). A map of the universe is shown in Fig.~\ref{fig:Illustration} C.
Movement is not for free. Traveling long-range connections costs more than using local ones. 
Travel can be fast but takes time; traversing the whole universe needs about three days. 
We define the distance between two sectors as one ``step'' (network distance 1) if they are connected by either a local or a long-range connection. 
For sectors that are not directly connected, we define their distance as their network (Dijkstra) distance.
The network of sectors has a diameter of 27 steps.
(See also \cite{Szell2012,PardusHP}.)

We denote the position of player $i$ at time $t$ (measured in days) by $x_i(t)$. 
Every position of a player is inside one sector and we define the distance of two players $i$ and $j$ as
\begin{equation}\label{eq:defR}
	r_{ij}(t) \equiv D(x_i(t), x_j(t) ) \quad ,
\end{equation}
where $D(x,y)$ is the Dijkstra distance between sectors $x$ and $y$. 
The players' (relative radial) velocity $v_{ij}(t)$, and  acceleration $a_{ij}$ are
\begin{eqnarray}
	\label{eq:defVelocity}
	v_{ij}(t) &\equiv& D(x_i(t+1), x_j(t+1)) - D(x_i(t), x_j(t) ) \quad , \nonumber \\
	a_{ij}(t) &\equiv& D(x_i(t+1), x_j(t+1)) + D(x_i(t-1), x_j(t-1) ) \nonumber \\ 
	&-& 2D(x_i(t), x_j(t) ) \quad ,
\end{eqnarray}
respectively. We use $N^{\beta}_{ij}(t)$ for the number of times $i$ interacted with $j$ in the time interval $[t, t+1[$.
$\beta=1,2,3$ specifies the type of interaction, communication, trade, and attack, respectively. 
For the case of no interaction we use $\beta=0$. 
For every interaction type $\beta$, the average distance $r^{\beta}$ is the conditional average over all distances between interacting players 
in the time window $[t, t+1[$,
\begin{equation}
 r^{\beta}= \langle r_{ij}(t) | i \,\, {\rm interacts \,\, with} \,\,  j {\rm \,\, through\,\, }  \beta {\rm \,\, in\,\, } [t, t+1[  \rangle_{ \{i,j,t\}  } .
 \end{equation} 
The average velocity $v^{\beta}$, and number of interactions $N^{\beta}$ are computed in the same way. 
The average acceleration as a function of the distance is calculated as the average over all interacting pairs $\{i,j\}$ given that $r_{ij}(t)=r$, 
\begin{eqnarray}
a^{\beta}(r) \equiv  &\langle & a_{ij}(t) | i \,\, {\rm interacts \,\, with}\,\,  j {\rm \,\, through\,\, }  \beta {\rm \,\, in\,\, } \nonumber \\  
  & [ &   t-1, t+1[,  \,\, {\rm and} \,\, r_{ij}(t)=r Ê \rangle_ { \{i,j,t\}} .
\end{eqnarray}
Negative velocity means motion toward each other.
Note that negative acceleration can mean three things: either that players increase speed towards each other, or that 
they slow down when moving apart, or that they change their direction from moving apart to moving towards each other.
Most players interact with several others at the same time and the effect of interactions between a pair of players 
is potentially disturbed by other interactions and factors. Assuming naive superposition of dyadic interactions by 
taking averages, random disturbances should cancel out.

\section{Results}

\begin{figure}
	\centering
	\includegraphics[width=0.4\textwidth]{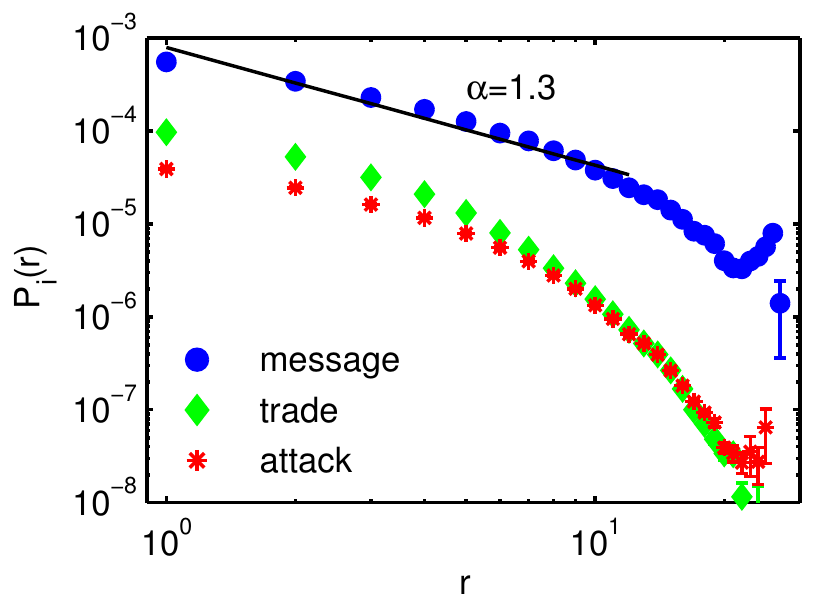}
	\caption{Probability of interactions of type $\beta$,  $P^{\beta}_i$ as a function of the distance $r$ between players.
	An approximate power exponent for messages is $\sim -1.3$ (black line).
	}
	\label{fig:Pr}
\end{figure}

\begin{table}
	\caption{Comparison of power exponents from the distance dependence of interaction probabilities for various data sets.}
\begin{tabular}{lcc}
		Interaction type & $\alpha$ & Source \\
		\hline
		Phone calls & 2 & \cite{Lambiotte2008} \\
		Call duration & 2 & \cite{Krings2009} \\
		Facebook, email, etc. & 1.98 & \cite{Levy2014} \\
		LiveJournal friendship & 1.2 & \cite{Liben-Nowell2005} \\
		Email & 1 & \cite{Adamic2005} \\
		Facebook friendship & 1.05 & \cite{Backstrom2010} \\
		Brightkite, Foursquare, Gowalla & 0.5--1 & \cite{Scellato2011} \\
		Gowalla friendship & 0.82 & \cite{Cho2011} \\
		Brightkite friendship & 0.83 & \cite{Cho2011} \\
		Twitter, Gowalla, Brightkite & 0.7 & \cite{Grabowicz2014} \\ 
		\hline
		Messages in \emph{Pardus} &  1.3 & Fig.~\ref{fig:Pr} \\
		\hline
	\end{tabular}
	\label{tab:exponents}
\end{table}

\subsection{Locality of interactions}
We define the probability $P^{\beta}_i (r)$ that $i$ interacts with any given $j$ that is a distance $r=r_{ij}(t)$ away from $i$, within a unit time interval $[t, t+1[$. 
$P^{\beta}_i (r)$ is shown as a function of $r$ in Fig.~\ref{fig:Pr}.
Even though the distributions are clearly not power-law, if an approximate power-law exponent was fitted for $\beta=$ messages, it yields $-1.3$. 
In Tab.~\ref{tab:exponents} this exponent is compared to those found in previous works. 
For trade and attack, the probability for an interaction decays faster than a power-law.
The stronger decay for trade or an attack can be explained by the fact that for these interactions players need to reduce their distance to  
zero within the 24h of observation. 
For $r>20$ we see a finite size effect. 

\begin{figure}
	\centering
	\includegraphics[width=0.4\textwidth]{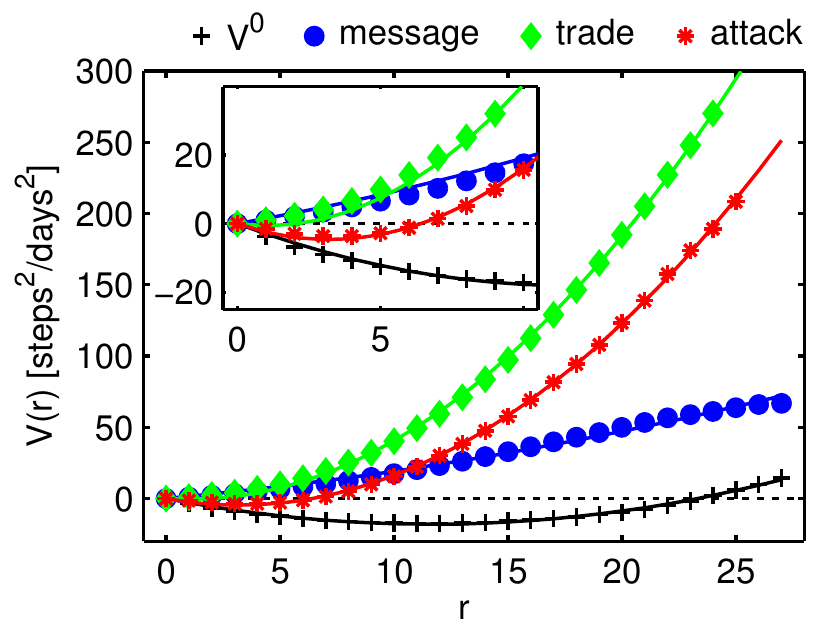}
	\caption{Interaction-specific potentials for messages, trade, and attacks, computed from Eq.~\eqref{eq:Vfromacc}.
	Solid lines are least-squares fits to a harmonic potential as defined in Eq.~\eqref{eq:HO}. 
	$V^0$ is a result of the background motion of non-interacting pairs of players. The inset is a magnification for small distances.
	 Clearly, the potential for attack shows a minimum at $r^{\rm attack}\sim3$. For all fits (lines) the explained variance is $R^2>0.99$.
	\label{fig:V_r}
	}
\end{figure}

\subsection{Interaction potentials of social interactions}
We assume $m=1$ in Eq.~\eqref{eq:NewtonRadial}, and that $V^0(r)$ is caused by the random 
background motion of players on the finite network of sectors.
The relative motion of players that do not interact with each other ($V_{}(r)=0$) is therefore governed by $V^0(r)$.
To estimate $V^0(r)$ in Eq.~\eqref{eq:NewtonRadial} we set $V_{}(r)=0$ and solve for 
$V^0(r) = \sum_{r^\prime=0}^r a^0(r^\prime) $, where $a^{\beta=0}$ means acceleration between non-interacting pairs.  
The interaction specific potential $V^{\beta}_{}(r)$ is  
\begin{equation}\label{eq:Vfromacc}
	V^{\beta}(r) = -\sum_{r^\prime=1}^r a^{\beta}(r^\prime) - a^0(r^\prime) \quad .
\end{equation}
Using $1$ step for $\D r$ in the integration, the resulting unit of $V$ is $\mathrm{steps}^2/\mathrm{days}^2$.
Starting the sum at $r^\prime=1$ sets the reference point $V^{\beta}(0) = 0$.

The resulting potentials are shown in Fig.~\ref{fig:V_r} for the three interaction types. 
They are well approximated by a harmonic and linear potential,
\begin{equation}
	\label{eq:HO}
	V^\beta(r) = \kappa^\beta r^2 - b^\beta r \quad , 
\end{equation}
where $\kappa^\beta$ is the respective ``force constant''.
The corresponding equilibrium distance is at $r_m^\beta = \frac {b^\beta} {2\kappa^\beta}$.
Potentials increase with distance without signs of saturation.  For communication this is consistent with the real-world observation 
that ``(...) the effect friends have on our movement grows with their distance from us'' \cite{Cho2011}.
For trades and attacks there is the simple explanation that players need to reduce their distance to zero at one instance so that the interaction is possible.

For non-interacting players ($V^0$) we find $\kappa^0 = 0.133 \; [0.130 ; 0.136]$, where the intervals give the 95\% confidence intervals of the fit. $b^0 = 3.1 \; [3.0 ; 3.2]$, corresponding to an equilibrium distance of $r_{m}^0 = 11.7$ 
steps, which is close to the average distance of any two randomly selected players (12.1 steps).
For trade we find  $\kappa^\mathrm{trade} = 0.52 \; [0.50 ; 0.53]$,  $b^\mathrm{trade} = 1.1 \; [0.9 ; 1.3]$, and  $r_{m}^\mathrm{trade} = 1.1$.
For attacks we get  $\kappa^\mathrm{attack} = 0.453 \; [0.447 ; 0.458]$, $b^\mathrm{attack} = 2.9 \; [2.8 ; 3.0]$, and $r_{m}^\mathrm{attack} = 3.2$.
$V^\mathrm{attack}$ is repulsive for $r<3$, which reflects  a common ``hit-and-run'' strategy of players.
Finally, for messages we have $\kappa^\mathrm{message} = 0.04 \; [0.03 ; 0.05]$, and $b^\mathrm{message} = -1.5 \; [-1.7 ; -1.3]$.
It is obvious from Fig.~\ref{fig:V_r} that $V^\mathrm{message}$ is mainly dominated by the linear term. 

\begin{figure}
	\includegraphics[width=0.5\textwidth]{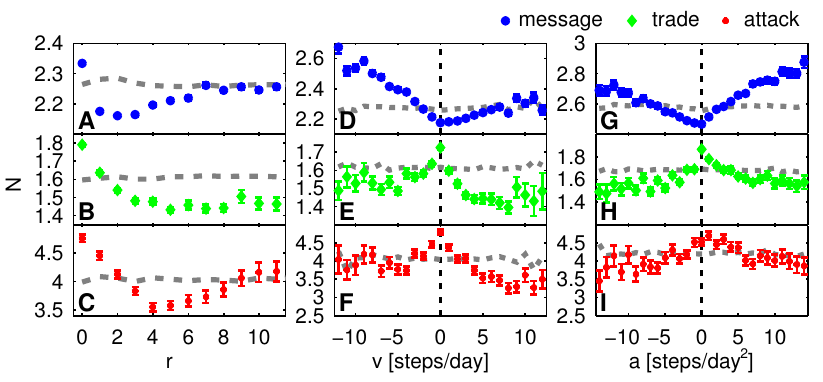}
	\caption{Average number of (uni-directional) interactions per day as a function of $r^{\beta}$, $v^{\beta}$, and $a^{\beta}$.
	 Errorbars are standard deviations of the mean. The results for randomized data are shown as dashed gray lines.
	 Clearly, interaction intensity is strongest for players in the same sector.
	 For trade and attack, stationary players (with velocity and acceleration zero) interact the most.
	 For messages, interaction is most intense for players who move towards each other but slow down.
	 }
	\label{fig:N_all}
\end{figure}

\subsection{Interaction strength and relative motion}
We analyze the average number $N^{\beta}$ of messages sent, or trades performed, or attacks carried out between players. 
Only pairs of interacting players are taken into account.
Figure \ref{fig:N_all}, panels A B C show the number of interactions $N^{\beta}$ as a function of distance $r^{\beta}$, respectively.
The gray lines indicate the level obtained from shuffled data (Methods) which serves as a baseline level. 
In A we see that the number of exchanged messages is strongly over-represented (above baseline) at zero distance. 
For all other distances messages are under-represented, reaching the baseline level for large distances. The minimum of messages is 
found at a distance of 2. Trades are over-represented for distances 0 and 1 (B); attacks for distances 0, 1, and 2 (C). For attacks a 
clear minimum is reached at distance 4. 
Panels D E F show the number of interactions $N^{\beta}$ as a function of velocity $v^{\beta}$. 
It is clear from D that the higher the velocity toward each other the higher is the number of messages. 
Small positive velocities (away from each other) are slightly underrepresented and approach the baseline for large velocities. 
The situation is different for trade and attack. Both positive and negative velocities are under-represented (positive ones slightly more). 
For trade (E) only zero velocity is over-represented. For attacks (F) absolute values of zero and 1 are over-represented.
This clearly shows the needed stationarity for trade and attack. 
Finally, panels G H I display the number of interactions $N^{\beta}$ as a function of acceleration $a^{\beta}$. 
G shows that the number of sent messages increases from a minimum at $a=0$ with the radial acceleration:
the highest number of messages is exchanged between players who move towards each other, but slow down.
For trades and attacks, accelerations close to zero are over-represented, while the baseline is approached for large accelerations.
Both are skewed towards positive values of the acceleration.

\begin{figure}
	\centering
	\includegraphics[width=0.5\textwidth]{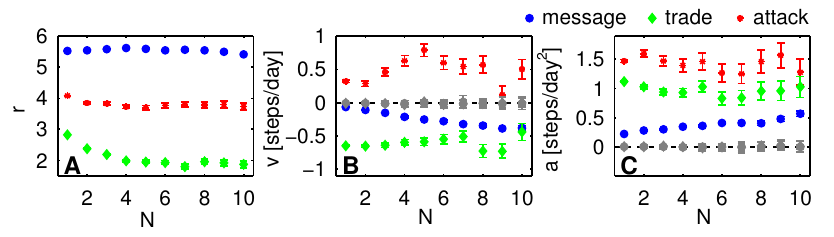}
	\caption{Average $r^{\beta}$, $v^{\beta}$, and $a^{\beta}$ as a function of the number of interactions $N^{\beta}$. 
	Errorbars are standard deviations of the mean. Randomized data are shown with gray symbols. 
	For the randomized data $r^{\beta} \sim 12.1$ for all interaction types and independent of the interaction strength (not shown).
	Characteristic distances are relatively constant in $N^{\beta}$, the characteristic convergence speed for communication increases with the number 
	of messages. The separation speed for attacks increases up to $N^{\rm attack}\sim 4$ and remains constant afterward. 
	Acceleration in positive direction, i.e. slowing of convergence for messages increases with $N^{\rm message}$.
	}
	\label{fig:all_N}
\end{figure}

In Fig.~\ref{fig:all_N} we show distance, velocity and  acceleration as a function of the number of interactions. 
Randomized data is shown by gray symbols. 
In panel A it is visible that the average distance for messages is about 6.5, for trade about 2, and 
just below 4 for attacks. These characteristic distances depend relatively little on the number of interactions $N^{\beta}$. 
The relative large distance for attacks might reflect a ``safety'' distance. 
For the randomized data the characteristic distance is $r \sim 12.1$ for all interaction types and independent of the number of interactions.
In Fig.~\ref{fig:all_N} B it is seen that for trade there is a typical constant convergence speed of $\sim0.6$ irrespective of interaction counts.
For messages the characteristic convergence speed increases slightly with the number of interactions. Very pronounced is the 
positive characteristic separation speed of attacks, which increases with the number of attacks until a plateau is reached from 
$N^{\rm attack}\sim4$ on.
Characteristic acceleration values for trade and attack are constant in $N^{\beta}$, for communication there is a slight increase 
of acceleration with $N^{\rm message}$.

\subsection{Relative motion before and after interactions}
\begin{figure}
	\centering
	\includegraphics[width=0.5\textwidth]{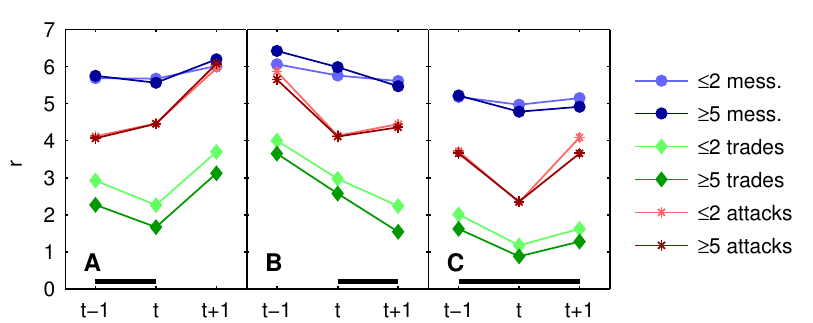}
	\caption{Relative distance $r^{\beta}$ of players before and after interactions.
	The black line denotes the time interval where interactions take place.
	A the players interact on the first day, but not on the second,
	B the players interact on the second day, but not on the first, 
	C the players interact on both days.
	}
	\label{fig:trajectories}
\end{figure}

Finally we study the effects on characteristic distances right before and after interaction events.  
In Fig.~\ref{fig:trajectories} we look at characteristic distances $r^{\beta}$ at three consecutive time points. 
Time windows where interactions happen are indicated by a black bar. 
Panel \ref{fig:trajectories} A shows how distances change after a period where interaction(s) occurred. 
It is clearly visible that when players cease to interact $r^{\beta}$ immediately increases (at $t+1$), i.e. they move away from each other.
The effect is especially pronounced for $\beta=$ trade and attack. 
There is a slight indication that the higher the number of trades, the closer the interacting players are (light and dark colors), see also Fig.~\ref{fig:all_N}.
From Fig.~\ref{fig:trajectories} B we learn that right before interactions take place, players approach each other (from $t-1$ to $t$). 
This effect can be understood in the following way:
If two players are closer to each other than the expected distance for a random pair of players, 
and assuming that they move independently and randomly, 
there is an increaed likelihood that they will be farther apart on the next day.
Similarly there is an increased likelihood that they have been farther apart one day earlier.
This effect constitutes the repulsion at $r<r_m^0$ captured by $V^0$.
Since $P_i(r)$ (Fig.~\ref{fig:Pr}) causes interacting players to be close to each other,
i.e. mostly closer than $r_m^0$, they move towards each other before an interaction.
The strongest effect is seen for attacks, for which the beginning separation after the attack is also visible at $t+1$.
For attacks we clearly identify a ``hit-and-run'' tactics where before the attack the players move towards each other (attacker closes in on victim). 
Right after the attack the attacker moves away from the victim, see panel A.
In Fig.~\ref{fig:trajectories} C we see that players who communicate with each other for a more extended period of time (2 days)
are closer than those who begin or end a communication, (compare to $r^{\rm message}$ in panels A and B). 
Their distance remains approximately constant over the three days. 
The same observations hold for trades. For attacks again the ``hit-and-run'' tactics is visible. 

\subsection{Model} 
We try to understand the observed interaction potential for messages by a simple model.  
The basic idea is to model a pair of random walkers $i$ and $j$ on the Pardus universe network Fig.~\ref{fig:Illustration} C. 
At every timestep $i$ sends a message to $j$ according to the distance-specific interaction probability $P^{\rm message}_i(r)$ (see Fig.~\ref{fig:Pr}).

Both players move to a randomly chosen new sector, which is $d$ steps from their current position. 
$d$ is sampled independently for each player from the empirical jump-distance distribution in the game, $\mathcal{P}(d)$ (Methods). 

If an interaction took place at the current timestep $t$, the player who initiated the interaction (sent the message) moves one step 
towards the other player along (one of) the shortest path(s). 
If an interaction took place at the previous timestep $t-1$, the player who initiated the interaction moves two steps 
towards the other player along (one of) the shortest path(s).
If both players have initiated an interaction, both move towards each other independently, 
each one step at the timestep he initiated the interaction and two steps in the timestep after the interaction.
Note that it is highly unlikely that both players initiate an interaction at the same time, since $P^{\rm message}_i(r)<10^{-3}$ for all $r$.

If $i$ and $j$ are already in the same sector, they remain there.
The procedure is repeated for $5\times 10^8$ days for 20 different random initializations. 

\begin{figure}
	\centering
	\includegraphics[width=0.4\textwidth]{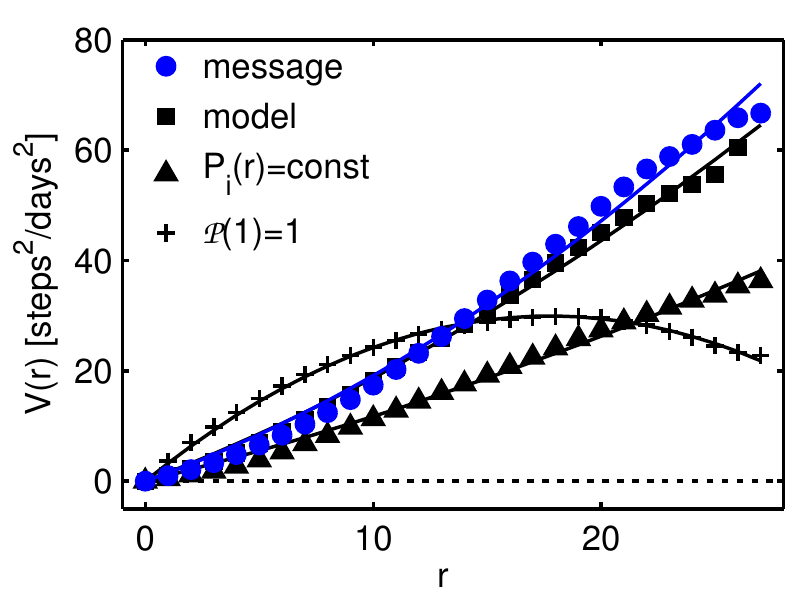}
	\caption{The interaction potential derived from the random walker model (squares) reproduces to a large extent the empirical potential for 
	messages (blue circles). The model uses the actual topology of the universe and assumes an attractive acceleration independent of the distance.
	Triangles show the situation where the distance-dependent interaction probabilities are set constant, pluses represent an unrealistic 
	constant jump-distance distribution of one step per time step. 
	Lines are least-squares fits to a harmonic potential as in Eq. \eqref{eq:HO}.
	}
	\label{fig:model}
\end{figure}

From the resulting relative movements of the players we derive the potential as above using Eq.~\eqref{eq:Vfromacc}.  
In Fig. \ref{fig:model} we see that the model (squares) reproduces the potential to a large extent well. 
The model is further able to explain the motion of players toward each other before an interaction that was mentioned in the context of Fig.~\ref{fig:trajectories} B. 
Note that the inputs were the topology of the universe, the empirical jump-distance distribution of players, the empirical distance-dependent interaction probability, and an acceleration
that does not depend on the distance.

To illustrate the importance of the distance-dependent interaction probability we set $P_i(r)={\rm const}$, and arrive at a potential that 
underestimates the empirical one (triangles). The effect of the jump-distance distribution is seen when we set $\mathcal{P}(1)=1$, 
i.e. players always move one step, bigger moves as well as no moves are forbidden. 
The resulting potential (pluses) has no more explanatory power. 
The same result is obtained when setting  $P_i(r)={\rm const}$, and $\mathcal{P}(1)=1$ (not shown). 
Note that for the case $\mathcal{P}(1)=1$, $V^0$ in Eq.~\eqref{eq:NewtonRadial} is no longer the one shown in Fig.~\ref{fig:V_r}.

\section{Discussion}
We presented a phenomenological study of a comprehensive data set of all interactions and all trajectories of the inhabitants of the virtual world \emph{Pardus}.
The most important finding in this work is that social interactions (mediated through exchange processes) lead to a measurable reduction or increase of 
relative distances and acceleration between people. The type of social interaction is essential for the details of these ``forces''.
Social exchange intensities are found highest for characters in the same place.
We find that the probability to interact decays with the distance (very) roughly as a power-law with exponent 1.3.
We relate this exponent to previous estimates of power-law exponents in a variety of real world settings \cite{Lambiotte2008,Krings2009,Levy2014,Liben-Nowell2005,Adamic2005,Backstrom2010,Cho2011,Scellato2011,Grabowicz2014}.

We find that relative movement patterns of people in the game can be well understood by approximately harmonic interaction potentials. 
These potentials vary considerably for different types of social interactions. 
$V^\mathrm{attack}$ has a clear minimum at distance of 3.2 steps, which is close to the observed average distance between players attacking 
each other (slightly below 4).
We observed a characteristic distance of players interacting through messages between 5 and 6 steps, and about 2 for those interacting through trade.
Since we can not define ``kinetic energy'' in a meaningful way in the game it is not possible to understand the typical distances as 
bound states of interacting individuals. 

Players who write messages to each other or trade with each other move towards each other; 
when they stop communicating, they move away from each other, comparable to the breaking of a bond.
This is consistent with earlier studies on real-world data, which showed that positive social ties are attractive 
\cite{Cho2011,Phithakkitnukoon2012}, and that the attraction increases with distance \cite{Cho2011}.
Players show a tendency to move towards each other shortly before they interact. 
The more intensive the communication (number of exchanged messages), the stronger is this effect.
For an attack, the attacker usually closes in on her victim before an attack and backs up directly after it. 
This ``hit-and-run'' tactics is clearly seen in the respective potential. 

The potential for the message exchange type can be understood by a simple model of random walkers, whose 
interactions cause an attractive acceleration. The model allows to disentangle several relevant effects such as the 
distance-dependence of interaction probability and the jump-distance distribution of walkers. Both are essential for explaining the observed 
potential. 
The main message of the model is that it is possible to understand the distance-dependence of inter-human forces from the 
distance-dependence of interaction probabilities. The acceleration following individual interactions can be uniform and independent 
of relative distance.

\section{Methods and Data}
We study data from one of three game universes of \emph{Pardus}, namely \emph{Artemis}.
Every day at 05:32 GMT a snapshot of all players' positions is taken for 1,238 consecutive days starting June 12, 2007. 
Messages, trades, and attacks are recorded with a time resolution of one second.
To accurately describe the motion of non-interacting players we exclude inactive players on a day-to-day basis:
For all results containing acceleration $a$ we only consider players who have performed at least one action between $t-1$ and $t+1$;
for all other results players must have at least performed one action between $t$ and $t+1$.
Given these requirements we get 3,414,091 data points of 31,496 unique players on 1,237 days. 
For the one-day time interval we have 3,126,842 occurrences of messages, 358,825 occurrences of trades, and 169,227 occurrences of attacks.
For all results regarding $a$, pairs of players who interacted only on one of two consecutive days are wrongly 
treated as two independent data points by the procedure described above.
To correct for this error the standard error of the mean is multiplied by a factor of 
$\sqrt{2}$ to account for the (at most) two-fold over-estimation of the number of independent data points.

\subsection{Randomized data}
Data are reshuffled  by assigning the positions ($x_i(t-1)$,) $x_i(t)$ and $x_i(t+1)$ to some other active player $j$ at random.
This is done separately for every day. This way individual trajectories, population densities, and interaction networks are left intact 
while the relation between positions and interactions are randomized.

\subsection{Jump-distance distribution $\mathcal{P}(d)$}
For all players $i$ who have at least performed one action between $t$ and $t+1$, we measure the jump-distance $d_i(t) \equiv D(x_i(t), x_i(t+1))$.
$\hat{\mathcal{P}}\left(D(x_i(t), x_i(t+1))=d\right)$ denotes the empirical probability distribution of these jump-distances.
The distribution used in the model, $\mathcal{P}(d)$, is derived from the measured probability of travel distances $\hat{\mathcal{P}}(d)$ by setting $\mathcal{P}(d)\equiv 0$ for $d>d^*$, where $d^*\equiv \min_x\max_yD(x,y)=20$, and normalizing.

\section{Acknowledgements}
We would like to thank B. Corominas-Murtra for helpful discussions.
We acknowledge support from the Austrian Science Fund FWF P23378, 
and the EU FP7 projects LASAGNE No. 318132 and MULTIPLEX No. 317532.

\end{document}